\documentclass[12pt,a4paper]{article}
\usepackage[latin1]{inputenc}
\usepackage{authblk}    
\usepackage{makeidx}    
\usepackage{graphicx}        
\usepackage{amsmath}
\usepackage{amsthm}
\usepackage{multicol}
\usepackage{multirow}
\usepackage{booktabs}
\usepackage{setspace}
\usepackage{hyperref}
\usepackage{tabularray}
\usepackage{xcolor}
\usepackage{subcaption}
\usepackage{geometry}
\usepackage{longtable}
\usepackage{yhmath}
\usepackage{diagbox}
\usepackage[natbibapa,nodoi]{apacite}
\newcommand{\mailadd}{\href{mailto:me@somewhere.com}{\vspace*{0.0cm}}} 

\providecommand{\keywords}[1]
{
   \small	
  \hspace{1cm} \textbf{\textit{Keywords:}} #1
}

 \makeatletter
 \renewcommand{\section}{\@startsection
  {section}{1}{0pt}{24pt}{12pt}{\bf \large}}
 
 \makeatother

\title{\textbf{{Biclustering bipartite networks via extended Mixture of Latent Trait Analyzers}}}

\author[a]{Dalila Failli}
\author[a]{Maria Francesca Marino}
\author[b]{Francesca Martella}
\affil[a]{\small Dipartimento di Statistica, Informatica, Applicazioni, Universit\`{a} degli Studi di Firenze, Viale Morgagni 59 - 50134 Firenze; \mailadd{\texttt{dalila.failli@unifi.it}}, \mailadd{\texttt{mariafrancesca.marino@unifi.it}}}
\affil[b]{\small Dipartimento di Scienze Statistiche, Sapienza Universit\`{a} di Roma, Piazzale Aldo Moro, 5 - 00185 Roma; \mailadd{\texttt{francesca.martella@uniroma1.it}

}}

\date{}
\begin{document}
\maketitle

\begin{abstract}
In the context of network data, bipartite networks are of particular interest, as they provide a useful description of systems representing relationships between sending and receiving nodes. In this framework, we extend the Mixture of Latent Trait Analyzers (MLTA) to perform a joint clustering of sending and receiving nodes, as in the biclustering framework. In detail, sending nodes are partitioned into clusters (called components) via a finite mixture of latent trait models. In each component, receiving nodes are partitioned into clusters (called segments) by adopting a flexible and parsimonious specification of the linear predictor. Dependence between receiving nodes is modeled via a multidimensional latent trait, as in the original MLTA specification. The proposal also allows for the inclusion of concomitant variables in the latent layer of the model, with the aim of understanding how they influence component formation. To estimate model parameters, an EM-type algorithm based on a Gauss-Hermite approximation of intractable integrals is proposed. A simulation study is conducted to test the performance of the model in terms of clustering and parameters' recovery. The proposed model is applied to a bipartite network on pediatric patients possibly affected by appendicitis with the objective of identifying groups of patients (sending nodes) being similar with respect to subsets of clinical conditions (receiving nodes).\\
\end{abstract}

\section{Introduction}\label{sec:1}
\begin{figure}[ht!]
\centering
\includegraphics[scale=.7]{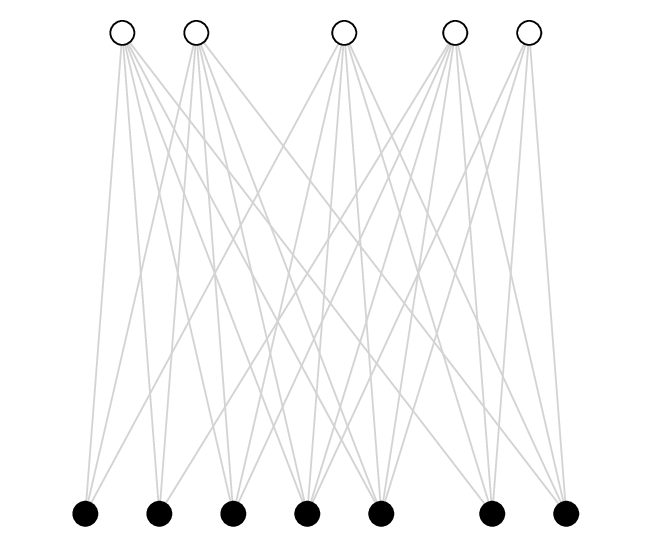}
\caption{Example of bipartite network.}
\label{fig:1}       
\end{figure}

Network data analysis is particularly relevant when the research interest lies in analyzing relationships between units. In this context, bipartite networks are gaining increasing importance. These represent connections between two disjoint sets of nodes, called sending and receiving nodes, respectively. Note that the terms ``sending" and ``receiving" are simply used to distinguish the two different sets, and do not directly refer to the direction of connections between nodes. The particular feature of this type of network is that connections are only allowed between nodes belonging to different sets, as shown in Figure \ref{fig:1}.
\\This peculiar type of network is useful for representing various phenomena, such as the boards of directors of companies \citep{davis2}, in which one group of nodes is represented by company directors, while the other group is represented by their boards. Another example in the economic context is given by the connections between countries and the products they export, which can be used to measure the complexity of a country economy \citep{hidalgo}. On the other hand, an example in the biological field is given by the bipartite network of metabolic reactions, where the two sets of nodes represent metabolites and reactions to which they participates, and by the protein-protein interaction network, representing interactions between proteins that form the so-called protein complexes \citep{newman}. In addition, researchers may be interested in analyzing the links between genes and protein products to identify some disease-specific susceptibility determinants \citep{bio}. Last, in the biomedical field, bipartite networks make it possible to represent potential drug-drug interactions, with the aim of analyzing or predicting any unknown interaction \citep{guimera}, while the disease-disease network shows how related diseases connect with each other \citep{barbasi}. Bipartite networks are also widely applied in social and behavioral research, with the aim of analyzing the participation of individuals to a set of social events \citep{davis}, or the behaviors adopted by patients to avoid infection during a pandemic \citep{failli2}.\\In network data analysis, a relevant research objective frequently concerns the identification of clusters of nodes sharing similar characteristics. This goal is frequently achieved by means of finite mixture specifications \citep{peel}. Firstly introduced by \cite{holland} and subsequently extended by, e.g., \cite{sniders}, \cite{nowiki}, \cite{daudin}, \cite{matias}, \cite{bart2018}, \cite{pandolfi}, Stochastic Block Models (SBMs) represent a particular type of finite mixture allowing to identify clusters of nodes sharing similar relational profiles. The latent position cluster model introduced by \cite{handcock} represents a further alternative, by allowing to identify communities of strongly connected nodes. This is done by considering a finite mixture of node-specific latent positions such that the probability of a tie between pair of nodes decreases as the distance between positions increases. Furthermore, \cite{vu} employs a finite mixture approach for clustering discrete-valued networks using scalable Exponential Family models. In the specific context of bipartite networks, \cite{aitkin2014} and \cite{aitkin2017} considered a latent class model to identify groups of sending nodes sharing similar (unobserved) features.\\However, since bipartite networks are characterized by separate sets of nodes, the research interest may also entail the joint clustering of sending and receiving nodes, as it is frequently done in data matrices where rows (units) and columns (variables) have interchangeable meaning and there is a meaningful relationship between them. In detail, one may be interested in looking for sending nodes that connect similarly to subsets of receiving nodes. This approach is commonly known in the literature as biclustering, co-clustering, block clustering, simultaneous clustering, or block modeling. Also in this framework, finite mixtures play a relevant role. For instance, \cite{govaert2003} and \cite{keribin} exploited a block mixture model to perform a simultaneous clustering of objects and variables through a mixture approach, while \cite{martella2008} and \cite{martella2017} proposed an extension of the mixture of factor analyzers (MFA) model for the simultaneous clustering of genes and tissues in microarray data. \cite{alfo} further extended these latter approaches to deal with longitudinal data. In the context of bipartite networks, \cite{wyse} and \cite{wyse2} employed the latent blockmodel (LBM) introduced by \cite{govaert2008} to perform a simultaneous clustering of the two sets of nodes in a bipartite network.\\In this paper, we extend the Mixture of Latent Trait Analyzers (MLTA) model, introduced by \cite{gollini2014} and \cite{gollini2020}. This originally combines features of latent class models and latent trait models, with the twofold objective of clustering sending nodes via a finite mixture specification and modeling the residual dependence between receiving nodes via a continuous multidimensional latent trait. Our proposal is to modify the MLTA in two ways. First, by allowing for a joint clustering of sending and receiving nodes: the former are partitioned into subsets called components and, in each of them, the latter are partitioned into subsets called segments. This goal is achieved by considering a finite mixture of latent trait analyzers providing a partitioning of sending nodes as in the original MLTA for bipartite networks. Then, within components, partitioning of receiving nodes is obtained by following an approach similar to that detailed by \cite{martella2017} and \cite{alfo}, based on a flexible and parsimonious specification of the linear predictor. Furthermore, following the approach by \cite{failli}, we allow for the inclusion of concomitant variables on the latent layer of the model to detect how they influence component formation.\\The performance of the model in terms of parameters' recovery and clustering is evaluated through a large scale simulation study based on a different number of nodes and partitions. Furthermore, the proposed approach is applied to a bipartite network of pediatric patients with potential appendicitis \citep{data}. Patients correspond to sending nodes, while their clinical conditions correspond to the receiving ones. A link between the two does exist if the given patient manifests the given clinical condition. The aim is twofold: 1) identifying groups of patients sharing specific subsets of clinical conditions; 2) taking into account how patients' characteristics influence the probability of belonging to a particular component.\\The paper is organized as follows. In Section \ref{sec:2}, we briefly describe the assumptions underlying the original specification of the MLTA model for bipartite networks. In Section \ref{sec:3}, we extend the MLTA model in a biclustering perspective, also describing parameter estimation and model selection. Section \ref{sec:4} shows the results of a simulation study conducted to verify the efficacy of the proposed approach in terms of parameters' recovery and clustering. Section \ref{sec:5} presents the application of the proposed model to the patient-condition bipartite network. Section \ref{sec:6} contains concluding remarks and details further extensions of the approach.

\section{MLTA model}\label{sec:2}
Let $Y_{ik}$ denote a binary random variable and $y_{ik}$ the corresponding
observed value for the $i$-th sending node, $i = 1, \dots, N$, and the $k$-th receiving node, $k=1,\dots,R$. A bipartite network can be formally described by a random incidence matrix $\boldsymbol{Y} = \{Y_{ik}\}$, with elements
\begin{equation*} 
    Y_{ik}=\Bigg\{\begin{array}{@{}l@{}}
    1 \quad \mbox{ if sending node } i \mbox{ is connected with receiving node } k,\\
    0 \quad \mbox{ otherwise.}
  \end{array}\end{equation*}
The MLTA model (\citealp{gollini2014, gollini2020}) combines latent class and latent trait analysis \citep{bart} by assuming that the set of $N$ sending nodes can be divided into $G$ distinct components and that the propensity for a sending node to be connected with a given receiving node depends also on a multidimensional, sending-specific, continuous, latent trait. Formally, it is assumed that every sending node belongs to one of $G$ unobserved components identified by a discrete latent variable $\boldsymbol{z}_i=(z_{i1}, \dots, z_{iG})^{\prime} \stackrel{iid}\sim \mbox{Multinomial}(1,(\eta_{1}, \dots, \eta_{G}))$, whose generic element is
 \begin{equation*}
    z_{ig}=\Bigg\{\begin{array}{@{}l@{}}
    1 \quad \mbox{ if sending node } i \mbox{ belongs to component } g,\\
    0 \quad \mbox{ otherwise.}
    \end{array}
\end{equation*}
The parameter $\eta_g$ denotes the probability that a randomly selected sending node belongs to component $g$, under the constraints that $\sum_{g=1}^G\eta_g=1$ and $\eta_g\geq0$, $g = 1, \dots ,G$.\\Furthermore, the model assumes the existence of a $D$-dimensional, sending-specific, continuous, latent trait $\boldsymbol{u}_i$ distributed according to a $D$-variate Gaussian density, with null mean vector and identity covariance matrix, i.e. $\boldsymbol{u}_i \sim \mathcal{N}_D(\boldsymbol{0}, \boldsymbol{I})$. This is assumed to capture the heterogeneity of sending nodes in the way they connect to receiving nodes. A practical example can be found in the network analyzed by \cite{failli2} and referenced in the previous section, where the sending nodes-specific latent trait can be interpreted as the unobserved propensity of each individual (sending node) to adopt specific measures (receiving nodes) to prevent COVID-19 infection. Also note that constraints on the distribution of $\boldsymbol{u}_i$ are required for identifiability purposes \citep{bart}.\\A local independence assumption is considered; that is, conditional on $\boldsymbol{z}_i$ and $\boldsymbol{u}_i$, response variables in the vector $\boldsymbol{Y}_i=(Y_{i1},\dots,Y_{iR})^{\prime}$ are assumed to be independent Bernoulli random variables with parameters $\pi_{gk}(\boldsymbol{u}_i)$, $k = 1, \dots ,R$, modeled through the following logistic function:
\begin{equation}
\label{pi}
    \pi_{gk}(\boldsymbol{u}_{i})=\text{Pr}(Y_{ik}=1\mid \boldsymbol{u}_{i}, z_{ig}=1)=\frac{1}{1+\exp[-(b_{gk}+\boldsymbol{w}_{gk}^{\prime}\boldsymbol{u}_{i})]}.
\end{equation}
Here, the model intercept $b_{gk}$ is a component-specific latent effect measuring the attractiveness of the $k$-th receiving node for those sending nodes belonging to the $g$-th component. A positive (respectively, low) value for this parameter indicates high (respectively, low) attractiveness of the $k$-th receiving node for sending nodes in the $g$-th component. Furthermore, the slope $\boldsymbol{w}_{gk}$ associated with the latent variable $\boldsymbol{u}_{i}$ is meant to capture the influence of the $D$-dimensional latent trait on the probability of a connection between the $k$-th receiving node and the sending nodes belonging to the $g$-th component. Statistically significant estimates of $\boldsymbol{w}_{gk}$ indicate association between receiving nodes, as well as the presence of heterogeneity of connections between nodes in the $g$-th component to the $k$-th receiving node, with respect to the baseline level provided by $b_{gk}$. Last, the discrete latent variable $\boldsymbol{z}_i$ and the continuous latent trait $\boldsymbol{u}_i$ are assumed to be independent.

\section{Extending the MLTA model for biclustering bipartite networks}\label{sec:3}
As it has been pointed out in Section \ref{sec:2}, our proposal aims first of all at performing a joint clustering of sending and receiving nodes. This is done by following an approach similar to that proposed by \cite{martella2017} and extended to the longitudinal framework by \cite{alfo}. In detail, the logistic function in Equation (\ref{pi}) is modified as:
\begin{equation}
\label{pi3}
    \pi_{gk}(\boldsymbol{u}_{i})=\text{Pr}(Y_{ik}=1\mid \boldsymbol{u}_{i}, z_{ig}=1)=\frac{1}{1+\exp[-({b_{g}}+{\boldsymbol{a}_{gk}^{\prime}}{(\boldsymbol{\mu}+\boldsymbol{u}_{i}))}]}.
\end{equation}
Here, the model intercept ${b_{g}}$ is a component-specific latent effect providing a baseline attractiveness measure for sending nodes in the $g$-th component, the parameter ${\boldsymbol{\mu}}$ is a $D$-dimensional vector of fixed effects which is assumed to be constant across components, and ${\boldsymbol{u}_{i}}\sim \mathcal{N}_D(\boldsymbol{0},\boldsymbol{I})$ is a $D$-dimensional, continuous, latent trait associated to sending node $i=1,\dots,N$.\\Last, ${\boldsymbol{a}_{gk}}$ is a $D$-dimensional row stochastic vector ($D\leq R$) with elements
\begin{equation*}
    a_{gkd}=\Bigg\{\begin{array}{@{}l@{}}
    1 \quad \mbox{ if, within the } g\mbox{-th component, receiving node } k \mbox{ belongs to segment } d,\\
    0 \quad \mbox{ otherwise.}
  \end{array}\end{equation*}
According to this definition, the vector $\boldsymbol{a}_{gk}$ allows to select a single term only from both $\boldsymbol{\mu}$ and $\boldsymbol{u}_i$. As far as this latter, it is important to notice that its dimension directly corresponds to the number of segments in which the $R$ receiving nodes are partitioned. Therefore, each latent trait $u_{id} \in \boldsymbol{u}_i$ allows to account for the heterogeneity of sending nodes in the way they connect to receiving nodes belonging to segment $d$. As before, constraints on the distribution of $\boldsymbol{u}_i$ are necessary for identifiability reasons. However, we can note that, under the proposed parametrization detailed in Equation (\ref{pi3}), we can also write $\boldsymbol{u}_i \sim \mathcal{N}_D(\boldsymbol{\mu},\boldsymbol{I})$ without affecting model identifiability. Based on these assumptions, for a sending node belonging to the $g$-th component and a receiving node belonging to the $d$-th segment, the connection probability would be 
\begin{equation*}
\label{pi4}
    \pi_{gk}({u}_{id})=\frac{1}{1+\exp[-({b_{g}}+{a_{gkd}({\mu}_{d}+{u}_{id}))}]},
\end{equation*}
so that $\mu_d$ is a fixed effect that increases or decreases the attractiveness of receiving nodes in the $d$-th segment for all sending nodes belonging to the $g$-th component with respect to the baseline level of the component, $b_g$. On the other side, $u_{id}$ is a unit- and segment- specific latent trait capturing the residual heterogeneity. When comparing Equation (\ref{pi3}) with Equation (\ref{pi}), we may notice that the intercept $b_{gk}$ in the former is constrained to be constant across receiving nodes in the latter; i.e., $b_{gk}=b_g$. This restriction is important from a biclustering perspective: in conjunction with the other parameters in the linear predictor it allows for a joint clustering of rows and columns of the incidence matrix.\\Following the strategy adopted by \cite{failli}, we also account for the effect that nodal attributes (i.e., observed characteristics of sending nodes) may have on component membership. This goal is achieved by letting the component probabilities $\eta_g$ vary across sending nodes via the following latent class regression model:
\begin{equation*}\label{eta}
    \eta(\boldsymbol{x}_i;\boldsymbol{\beta}_g)=\text{Pr}(z_{ig} = 1\mid \boldsymbol{x}_i; \boldsymbol{\beta}_g)=\frac{\exp\{\boldsymbol{x}^{\prime}_i\boldsymbol{\beta}_g\}}{1+\sum_{g'=2}^G\exp\{\boldsymbol{x}^{\prime}_i\boldsymbol{\beta}_{g'}\}}, \quad g=2, \dots, G.
\end{equation*}
Here, $\boldsymbol{\beta}_g$ denotes a $J$-dimensional vector of coefficients measuring the impact of nodal attributes $\boldsymbol{x}_i=(x_{i1},\dots,x_{iJ})^{\prime}$ on $\eta(\boldsymbol{x}_i;\boldsymbol{\beta}_g)$.

\subsection{Parameters estimation}
Let $\boldsymbol{\theta}=(\boldsymbol{\beta}_2, \dots, \boldsymbol{\beta}_G, b_{1}, \dots, b_{G}, \boldsymbol{a}_{11}, \dots, \boldsymbol{a}_{GR}, \mu_1, \dots, \mu_D)^{\prime}$ represent the vector of all free model parameters. Given the assumptions described in the previous section, the likelihood function of the model can be written as:
\begin{equation}
    L(\boldsymbol{\theta})=\prod_{i=1}^N L_i(\boldsymbol{\theta}) = \prod_{i=1}^N \Bigg(\sum_{g=1}^G \eta(\boldsymbol{x}_i;\boldsymbol{\beta}_g) \int_{{R}^D} f(\boldsymbol{y}_{i}\mid \boldsymbol{u}_i, z_{ig}=1)f(\boldsymbol{u}_i)d\boldsymbol{u}_i \Bigg),
    \label{loglikelihood}
\end{equation}
where $f(\boldsymbol{y}_{i} \mid \boldsymbol{u}_i,z_{ig}=1)=\prod_{k=1}^R f({y}_{ik} \mid \boldsymbol{u}_i,z_{ig}=1)=\prod_{k=1}^R(\pi_{gk}(\boldsymbol{u}_i))^{y_{ik}}(1-\pi_{gk}(\boldsymbol{u}_i))^{1-y_{ik}}$ denotes the product of Bernoulli probability mass functions. Note that, as standard with GLMMs, the multi-dimensional integral in Equation (\ref{loglikelihood}) cannot be computed analytically. Maximum likelihood estimates of model parameters can be obtained by maximizing the log-likelihood function above via an EM-type algorithm \citep{dempster}; this represents a standard choice when dealing with latent variables. The EM algorithm alternates two steps until convergence: at the E-step, the expected value of the complete data log-likelihood, given the observed data and the current parameter values, is computed; at the M-step, the expected complete data log-likelihood is maximized with respect to model parameters.\\In this framework, the complete data log-likelihood function is defined as
\begin{equation}\label{complete}
    \ell_c(\boldsymbol{\theta})=\sum_{i=1}^N \sum_{g=1}^G z_{ig}\log [f(\boldsymbol{y}_i\mid \boldsymbol{u}_i, z_{ig}=1)]+\sum_{i=1}^N\sum_{g=1}^G z_{ig}\log [\eta(\boldsymbol{x}_i;\boldsymbol{\beta}_g)] +\sum_{i=1}^N\log [f(\boldsymbol{u}_i)].
\end{equation}
As pointed out before, at the $(t+1)$-th iteration of the algorithm, the E-step consists in computing the expectation of Equation (\ref{complete}), conditional on the observed data and the current parameter estimates $\boldsymbol{\theta}^{(t)}$; that is
\begin{align}\label{Q}
    Q(\boldsymbol{\theta}\mid \boldsymbol{\theta}^{(t)})=& \sum_{i=1}^N \sum_{g=1}^G \int_{R^D} f(z_{ig}=1, \boldsymbol{u}_i\mid \boldsymbol{y}_i;\boldsymbol{\theta}^{(t)}) \log [ f(\boldsymbol{y}_i\mid \boldsymbol{u}_i, z_{ig}=1)]\; d\boldsymbol{u}_i \; + \nonumber \\
    &+ \sum_{i=1}^N \sum_{g=1}^G \hat{z}_{ig}^{(t+1)} \log [\eta(\boldsymbol{x}_i;\boldsymbol{\beta}_g^{(t)})] \; + \nonumber \\
    &+ \sum_{i=1}^N \int_{R^D} f(\boldsymbol{u}_i\mid \boldsymbol{y}_i; \boldsymbol{\theta}^{(t)}) \log [f(\boldsymbol{u}_i)]\; d\boldsymbol{u}_i.
\end{align}
Here, $f(z_{ig}=1, \boldsymbol{u}_i\mid \boldsymbol{y}_i;\boldsymbol{\theta}^{(t)})=\hat{z}_{ig}^{(t+1)}f(\boldsymbol{u}_i\mid z_{ig}=1,\boldsymbol{y}_i;\boldsymbol{\theta}^{(t)})$ denote the joint posterior density of $z_{ig}$ and $\boldsymbol{u}_i$, with $\hat{z}_{ig}^{(t+1)}$ being the posterior expectation of $z_{ig}$, given the current parameter estimates, $\boldsymbol{\theta}^{(t)}$, and the observed data, $\boldsymbol{y}_i$. Removing the dependence on the iteration index and doing a little algebra, these latter quantities can be computed as
\begin{align}\label{z}
    \hat{z}_{ig} = \frac{\eta(\boldsymbol{x}_i;\boldsymbol{\beta}_g)\int_{R^D} {f}(\boldsymbol{y}_i\mid \boldsymbol{u}_i, z_{ig}=1)f(\boldsymbol{u}_i) d\boldsymbol{u}_i}{\sum_{g=1}^G \eta(\boldsymbol{x}_i;\boldsymbol{\beta}_g)\int_{R^D} {f}(\boldsymbol{y}_i\mid \boldsymbol{u}_i, z_{ig}=1)f(\boldsymbol{u}_i) d\boldsymbol{u}_i}.
\end{align}
The M-step of the algorithm consists in updating the model parameters by maximising the expected complete data log-likelihood function in Equation (\ref{Q}) with respect to $\boldsymbol{\theta}$. In detail, parameters $b_g$ and $\boldsymbol{\mu}$ are estimated by finding the zeros of the expected score function, computed with respect to the posterior distribution of the indicator variables, $z_{ig}$, and the continuous latent trait, $\boldsymbol{u}_i$. As far as the vector $\boldsymbol{a}_{gk}$ is concerned, this is estimated, for fixed $g$ and $k$, by identifying the segment $d$ providing the maximum value for the expected complete data log-likelihood. Finally, the multinomial logit coefficients, ${\boldsymbol{\beta}}_g$, are estimated by maximising the likelihood of a multinomial logit model via a Newton-Raphson algorithm, with weights provided by the $\hat{z}_{ig}$'s derived at the E-step of the algorithm, as in Equation (\ref{z}). Prior probabilities $\eta(\boldsymbol{x}_i;\boldsymbol{\beta}_g)$ are updated accordingly.\\The procedure is repeated until convergence, which occurs when 
$$\mid \mid \ell_c(\hat{\boldsymbol{\theta}}^{(t+1)})-\ell_c(\hat{\boldsymbol{\theta}}^{(t)})\mid \mid < \epsilon, $$
where $\epsilon>0$ denotes a given tolerance level. In the following, we set $\epsilon=10^{-4}$.

\subsubsection{Approximating intractable integrals}\label{gauss}
As it is clear from the previous section, the EM algorithm requires the calculation of multidimensional integrals that do not have closed-form solutions. In this context, techniques based on Gaussian quadrature approximations are rather frequent choices, see e.g. \cite{pinheiro} and \cite{skrondal}. Here, the main idea is to approximate an integral through a weighted sum of the integrand function, evaluated on a fixed set of abscissas with given weights.\\In our framework, based on Gaussianity of the terms $\boldsymbol{u}_i$, $i=1,\dots,N$, we may rely on a Gauss-Hermite quadrature, providing an approximation of the type
\begin{equation*}\label{form}
    \int_{R^D} h(\boldsymbol{u}_i)e^{-\mid\mid \boldsymbol{u}_i \mid \mid^{2}} d\boldsymbol{u}_i \approx \sum_{q_1 \dots q_D}h(\boldsymbol{u}_{q_1 \dots q_D})\prod_{l=1}^D \omega_{q_{l}},
\end{equation*}
where $$\omega_q=\frac{2^{Q-1}Q!\sqrt{\omega}}{Q^2[H_{Q-1}\omega_q]^2}.$$ Here, $\boldsymbol{u}_{q_1 \dots q_D}=(u_{q_1},\dots,u_{q_D})^{\prime}$ represents a $Q$-tuple defined by the Cartesian product of the roots of the $Q$-th order Gauss-Hermite polynomial $H_{Q}(\boldsymbol{u_i})$, for $q_l \in \{1, \dots, Q\}, \; \text{and}\; l \in \{1, \dots, D\}$. On the other side, $\omega_{q_{l}} $ are the corresponding quadrature weights, and $\sum_{q_1 \dots q_D}$ is used as a shorthand for $\sum_{q_1=1}^{Q}\dots\sum_{q_D=1}^{Q}$.\\Let's focus on solving integrals of the form 
\begin{equation}\label{zeta}
    \zeta_{ig}=\int_{R^D} f(\boldsymbol{y}_i\mid \boldsymbol{u}_i, z_{ig}=1) f(\boldsymbol{u}_i) d\boldsymbol{u}_i,
\end{equation}
required for deriving both the likelihood and the posterior probabilities $\hat{z}_{ig}$; the computation follows a similar root when the expected score functions to estimate $b_g$, $\boldsymbol{\mu}$, and $\boldsymbol{a}_{gk}$ need to be derived. As stated before, since the $D$-dimensional continuous latent trait $\boldsymbol{u}_i$ follows a $D$-variate standard Gaussian distribution, i.e. $f(\boldsymbol{u}_i) = (2\pi)^{-D/2}\exp\{-\frac{1}{2}\boldsymbol{u}_i^{\prime}\boldsymbol{u}_i\}$, the integral in (\ref{zeta}) can be rewritten as a function of variables $\tilde{\boldsymbol{u}}_i=\frac{\boldsymbol{u}_i}{\sqrt{2}}$; that is:
\begin{align}
   \zeta_{ig} &\propto \sqrt{2^D} \int_{R^D} {f}(\boldsymbol{y}_{i}\mid \sqrt{2}\tilde{\boldsymbol{u}}_i, z_{ig}=1)\;e^{-\mid\mid \tilde{\boldsymbol{u}}_i\mid\mid ^2}d\tilde{\boldsymbol{u}}_i, \nonumber
\end{align}
so that the following approximation directly holds:
\begin{align}
\zeta_{ig} &\approx \sqrt{2^D} \sum_{q_1 \dots q_D} {f}(\boldsymbol{y}_{i}\mid \boldsymbol{u}^{*}_{q_1 \dots q_D}, z_{ig}=1)\;\prod_{l=1}^D \omega_{q_{l}}, \nonumber
\end{align}
where $\boldsymbol{u}^{*}_{q_1 \dots q_D}=\sqrt{2}\boldsymbol{u}^{}_{q_1 \dots q_D}$.

\subsubsection{Parameters initialization and clustering}\label{init}
As it is typically the case with latent variables, an initialization strategy based on multiple starting points can help the estimation algorithm not to get trapped in local maxima of the likelihood function. In detail, the estimation algorithm can be run starting from different initial values, then retaining the solution providing the maximum (log)-likelihood value. In our proposal, the following initialization strategy is adopted. The $\boldsymbol{\beta}_g$ parameters, $g=1,\dots,G$, are all initially set equal to zero, while parameters $b_g$ and $\boldsymbol{\mu}$ are initialized via a random generation from a Gaussian distribution with zero mean and unit variance. Furthermore, the vector $\boldsymbol{a}_{gk}$ is initialized by applying a k-means clustering method on both rows and columns of the incidence matrix.\\
At convergence of the algorithm, each sending node can be assigned to the $g$-th component via a Maximum a Posteriori (MAP; \citealp{bishop}) rule on the basis of the estimated posterior probabilities $\hat{z}_{ig}$, while each receiving node is automatically assigned to one of the segments according to the vector $\hat{\boldsymbol{a}}_{gk}$ obtained at convergence of the EM algorithm.

\subsection{Standard errors and model selection}
Standard errors for the estimates obtained via the EM algorithm can be obtained following the method suggested by \cite{louis} and \cite{oakes}, and by considering a sandwich formula \citep{royall, white}. Denoting by $\hat{\boldsymbol{\theta}}$ the vector of parameter estimates, the corresponding standard errors are obtained as the square root of the diagonal elements of the following covariance matrix
$$\hat{Cov}(\hat{\boldsymbol{\theta}})=\hat{\boldsymbol{J}}(\hat{\boldsymbol{\theta}})^{-1}\hat{\boldsymbol{K}}(\hat{\boldsymbol{\theta}})\hat{\boldsymbol{J}}(\hat{\boldsymbol{\theta}})^{-1}.$$
Here, $\hat{\boldsymbol{J}}(\hat{\boldsymbol{\theta}})$ is the observed information matrix; its estimate is obtained by computing the first numerical derivative of the score vector $S(\boldsymbol{\theta})=\frac{\partial\ell(\boldsymbol{\theta})}{\partial\boldsymbol{\theta}}$, evaluated at $\hat{\boldsymbol{\theta}}$. Furthermore, $\hat{\boldsymbol{K}}(\hat{\boldsymbol{\theta}})$ provides an estimate for the covariance matrix of the score vector and is computed as $\hat{\boldsymbol{K}}(\hat{\boldsymbol{\theta}})=\sum_{i=1}^N S_i(\hat{\boldsymbol{\theta}})S(\hat{\boldsymbol{\theta}})^{\prime}$, where $S_i(\hat{\boldsymbol{\theta}})$ denotes the individual contribution to the score function for the $i$-th sending node, evaluated at $\hat{\boldsymbol{\theta}}$.\\As far as model selection is concerned, the number of components $G$ and the number of segments $D$ are considered as fixed quantities. The model is estimated for different values of $G$ and $D$, and, then the optimal solution is the one corresponding to the smallest value of the chosen information criterion, such as the {Bayesian Information Criterion} (BIC; \citealp{schwarz})
$$\text{BIC}=-2\ell(\hat{\boldsymbol{\theta}}) + \nu\log(N),$$
where $\nu$ is the number of free parameters, or the model for which the {Integrated Classification Likelihood (ICL; \citealp{icl})
$$ \text{ICL}=\text{BIC}-\sum_{g=1}^G\sum_{i=1}^N\hat{z}_{ig}\log(\hat{z}_{ig})$$
is maximum.

\section{Simulation study}
\label{sec:4}
The performance of the model in terms of parameters' recovery and clustering is evaluated through a simulation study based on a different number of nodes, components and segments, as described in Sections \ref{setup}-\ref{param}. 

\subsection{Simulation setup}\label{setup}
Twelve different scenarios with $100$ simulated samples are considered; these are based on a varying number of sending nodes ($N=100,\; N=500,\; N=1000$), receiving nodes ($R=20,\; R=30$),  components $(G=3,\;G=4)$, and segments $(D=2, \; D=3)$. As regards the latent class variable, we considered a single nodal attribute, ${x}_i$, drawn from a Gaussian distribution with mean and variance both equal to 1. The latent structure is also defined by setting $\boldsymbol{\beta}_2 = [1, -0.4]$ and $\boldsymbol{\beta}_3 = [1.5, -0.9]$, for $G = 3$, and $\boldsymbol{\beta}_2 = [1, -0.4]$, $\boldsymbol{\beta}_3 = [1.5, -0.9]$, and $\boldsymbol{\beta}_4 = [2, -1.3]$, for $G = 4$. Furthermore, we set $\boldsymbol{b}=[-1.7, 0, 1.7]$ and $\boldsymbol{\mu}=[-2,0.5]$, when $G=3$ and $D=2$, and $\boldsymbol{b}=[-1.7, 0, 1.7, 0.7]$ and $\boldsymbol{\mu}=[-2,0.5, 1.5]$, when $G=4$ and $D=3$. Furthermore, the multidimensional continuous latent trait is randomly drawn by a multivariate Gaussian distribution with zero mean vector and identity covariance matrix, i.e. $\boldsymbol{u}_i\sim \mathcal{N}_{D}(\boldsymbol{0}, \boldsymbol{I})$. In each scenario, a multi-start strategy based on $100$ random starts is adopted.\\
The choice of this set up is motivated by the need of studying the behavior of the proposed method when the network size increases and the biclustering structure is more complex. We expect that the ability of the method to recover the true partition and the true parameter values improves with network size, while the performance should be less satisfying with more complex structures.

\subsection{Simulation results: clustering recovery}
The ability of the proposal in correctly classifying sending and receiving nodes is evaluated through the Adjusted Rand Index (ARI; \citealp{rand}), which is a standard measures of the agreement between the true and estimated partitions. Its expected value is zero in the case of random partition, while it is 1 in the case of perfect agreement between the two partitions. Results of the simulation study are shown in Tables \ref{tab:1} and \ref{tab:2}. Looking at these tables, we note that, as the number of sending and receiving nodes increases, the classification improves for both sending and receiving nodes. However, ARI values worsen as the number of partitions increases, as it is evident when comparing Table \ref{tab:1} with Table \ref{tab:2}. In detail, when $G = 4$ and $D=3$, larger network sizes are needed to obtain good clustering performance. We expect the classification ability of the model to further improve when considering bipartite networks with a larger number of nodes, as in the real-data application discussed in Section \ref{sec:4}.

\begin{table}[ht!]
\caption{Adjusted Rand Index mean (median) across samples for $G = 3$, $D = 2$, and varying $N$ and $R$.}
\label{tab:1}       
\centering
\begin{tabular}{rrcc|cc}
\noalign{\smallskip}\hline\noalign{\smallskip}
& & \multicolumn{2}{c}{$R=20$} &  \multicolumn{2}{c}{$R=30$} \\
\cmidrule(lr){3-4} %
\cmidrule(lr){5-6}
& & Sending  &  Receiving & Sending  &  Receiving   \\
\noalign{\smallskip}\hline\noalign{\smallskip}
\multirow{3}{*}{\textit{N}}& 100 & 0.71 (0.72)  &  0.83 (0.93) & 0.82 (0.83) & 0.91 (0.96)\\ 
& 500 & 0.75 (0.75) & 0.96 (1.00) & 0.83 (0.83) & 1.00 (1.00)\\
& 1000 & 0.75 (0.77) & 0.96 (1.00) & 0.83 (0.83) & 1.00 (1.00)\\
\noalign{\smallskip}\hline\noalign{\smallskip}
\end{tabular}
\end{table}

\begin{table}[ht!]
\caption{Adjusted Rand Index mean (median) across samples for $G = 4$, $D = 3$, and varying $N$ and $R$.}
\label{tab:2}       
\centering
\begin{tabular}{rrcc|cc}
\noalign{\smallskip}\hline\noalign{\smallskip}
& & \multicolumn{2}{c}{$R=20$} &  \multicolumn{2}{c}{$R=30$} \\
\cmidrule(lr){3-4} %
\cmidrule(lr){5-6}
& & Sending  &  Receiving & Sending  & Receiving   \\
\noalign{\smallskip}\hline\noalign{\smallskip}
\multirow{3}{*}{\textit{N}}& 100 & 0.64 (0.65)  & 0.38 (0.40) & 0.78 (0.79) & 0.49 (0.53)\\ 
& 500 & 0.69 (0.71)  & 0.48 (0.48) & 0.83 (0.83) & 0.63 (0.67)\\
& 1000 &  0.70 (0.72) & 0.49 (0.51) & 0.84 (0.84) & 0.63 (0.68)\\
\noalign{\smallskip}\hline\noalign{\smallskip}
\end{tabular}
\end{table}

\subsection{Simulation results: parameters' recovery}\label{param}
Tables \ref{tab:block} and \ref{tab:block2} show the Mean Squared Error (MSE) values across samples for $\boldsymbol{b}=[b_1,\dots,b_G]^{\prime}$, $\boldsymbol{\mu}=[\mu_1,\dots,\mu_D]^{\prime}$, and $\boldsymbol{\beta}=[\boldsymbol{\beta}_2,\dots,\boldsymbol{\beta}_{G}]^{\prime}$, obtained when letting the number of nodes, components, and segments vary. The MSE is computed as the mean of the squared differences between the estimates and the true value of model parameters. Therefore, the smaller the MSE, the closer the estimates and the true parameters. Looking at Tables \ref{tab:block} and \ref{tab:block2}, it is evident that, as the size of the network increases, we are more and more able to identify the true values of model parameters. However, MSEs increase as the number of blocks increases. In detail, for $G = 4$ and $D=3$, larger network sizes are needed to obtain satisfying results. Again, we expect that by further increasing the number of sending and receiving nodes, estimates will improve.

\begin{table}[ht!]
     \caption{{MSE} values across samples for $\boldsymbol{b}$, $\boldsymbol{\mu}$ and $\boldsymbol{\beta}$ for $G=3$ and $D=2$, and varying $N$ and $R$.}   
\label{tab:block}       
\centering
\begin{tabular}{rrccc}  
\noalign{\smallskip}\hline\noalign{\smallskip}
 & & \multicolumn{3}{c}{$R=20$} \\
\cmidrule(lr){3-5} %
& & $\boldsymbol{b}$  &   $\boldsymbol{\mu}$ & $\boldsymbol{\beta}$\\
\noalign{\smallskip}\hline\noalign{\smallskip}
\multirow{3}{*}{\textit{N}}& 100 & $[0.25, 0, 0.09]$ & $[0.03, 0]$  & $[0.62, 0.43, 0.14, 0.15]$ \\ 
& 500 & $[0.05, 0, 0.01]$ & $[0.02, 0]$ & $[0.09, 0.07, 0.02, 0.02]$ \\
& 1000 & $[0.03, 0, 0.01]$ & $[0.02, 0]$ & $[0.08, 0.06, 0.01, 0.01]$\\
\noalign{\smallskip}\hline\noalign{\smallskip}
& & \multicolumn{3}{c}{$R=30$} \\
\cmidrule(lr){3-5} %
& & $\boldsymbol{b}$  &  $\boldsymbol{\mu}$  & $\boldsymbol{\beta}$ \\
\noalign{\smallskip}\hline\noalign{\smallskip}
\multirow{3}{*}{\textit{N}} & 100 & $[0.12, 0, 0.06]$ & $[0.03, 0]$ & $[0.34, 0.27, 0.11, 0.10]$ \\ 
& 500 & $[0.02, 0, 0.01]$  & $[0.01, 0]$ & $[0.05, 0.05, 0.02, 0.02]$\\
& 1000 & $[0.01, 0, 0.01]$  & $[0.00, 0]$ & $[0.02, 0.02, 0.01, 0.01]$ \\
\noalign{\smallskip}\hline\noalign{\smallskip}
\end{tabular}
\end{table}

\begin{table}[ht!]
     \caption{MSE values across samples for $\boldsymbol{b}$, $\boldsymbol{\mu}$ and $\boldsymbol{\beta}$ for $G=4$ and $D=3$, and varying $N$ and $R$.}   
\label{tab:block2}       
\centering
\begin{tabular}{rrccc}  
\noalign{\smallskip}\hline\noalign{\smallskip}
 & & \multicolumn{3}{c}{$R=20$} \\
\cmidrule(lr){3-5} %
& & $\boldsymbol{b}$  &   $\boldsymbol{\mu}$ & $\boldsymbol{\beta}$\\
\noalign{\smallskip}\hline\noalign{\smallskip}
\multirow{3}{*}{\textit{N}}& 100 & $[0.56, 0, 0.16, 0.30]$ & $[0.25, 0, 0.27]$  & $[1.54, 0.98, 1.03, 0.50, 0.39, 0.24]$ \\ 
& 500 & $[0.27, 0, 0.07, 0.17]$ & $[0.22, 0, 0.40]$ & $[0.20, 0.17, 0.19, 0.04, 0.05, 0.05]$ \\
& 1000 &  $[0.13, 0, 0.01, 0.04]$ & $[0.08, 0, 0.38]$  & $[0.12, 0.07, 0.06, 0.02, 0.02, 0.01]$ \\
\noalign{\smallskip}\hline\noalign{\smallskip}
& & \multicolumn{3}{c}{$R=30$}\\
\cmidrule(lr){3-5} %
& & $\boldsymbol{b}$  &   $\boldsymbol{\mu}$ & $\boldsymbol{\beta}$\\
\noalign{\smallskip}\hline\noalign{\smallskip}
\multirow{3}{*}{\textit{N}}& 100  & $[0.32, 0, 0.11, 0.21]$ & $[0.12, 0, 0.15]$ & $[0.81, 0.60, 0.52, 0.23, 0.25, 0.20]$ \\ 
& 500 & $[0.11, 0, 0.07, 0.11]$  & $[0.08, 0, 0.18]$ & $[0.09, 0.08, 0.08, 0.03, 0.04, 0.03]$\\
& 1000 & $[0.10, 0, 0.06, 0.10]$ & $[0.03, 0, 0.17]$ & $[0.06, 0.06, 0.03, 0.03, 0.04, 0.01]$ \\
\noalign{\smallskip}\hline\noalign{\smallskip}
\end{tabular}
\end{table}

\section{Application to the pediatric patients bipartite network}
\label{sec:5}
In this section, we analyze the bipartite network entailing connections between pediatric patients with suspected appendicitis and their clinical conditions \citep{data}. In detail, in Section \ref{description}, we describe the data set, while in Section \ref{analysis} we analyze the data via the proposed modeling approach.

\subsection{The network}\label{description}
The bipartite network considered in the analysis is built starting from a data set that collects information from a cohort of 782 pediatric patients with abdominal pain and suspected appendicitis, admitted to St. Hedwig Children's Hospital in Regensburg (Germany), between 2016 and 2021. For each patient, we have information on the results of ultrasound (US) images, laboratory tests, and clinical scores, for a total of 54 clinical conditions. By excluding missing data, we reduce the data matrix to 542 rows (the sending nodes/patients) and 23 columns (the receiving nodes/clinical conditions). Table \ref{descr_var} shows a detailed description of such conditions.
\begin{table}[ht!]
\tiny
    \centering
    \begin{tabular}{l p{10cm}}
    \hline
Alvarado Score & Score based on symptoms, signs, and laboratory test. A score between 0 and 3 corresponds to unlikely appendicitis, 4 to 6 to possible appendicitis, and a score between 7 and 10 to very likely/certain appendicitis.\\              
Pediatric Appendicitis Score & Score based on patient's history and examination. A score of five or less rules out appendicitis, while a score of six or more makes a true case of appendicitis highly probable.\\ 
Appendix on US & Detectability of the vermiform appendix during sonographic examination. The variable assume value 1 if detected and 0 otherwise.\\                    
Migratory Pain & Abdominal pain from epigastrium to the right lower quadrant.\\                      
Lower Right Abdominal Pain & Right iliac fossa pain detected on palpation.\\       
Contralateral Rebound Tenderness & Pain of the contralateral side during the release of pressure over the abdomen.\\
Coughing Pain  & Abdominal pain by forced cough.\\                   
Nausea          & Feeling of sickness/ejection of contents from stomach through the mouth.\\                 
Loss of Appetite            &    Lack or reduction of appetite.\\    
Body Temperature & Temperature measured by a thermometer placed in the rectum or in the auditory canal.\\                    
WBC Count               &    Number of leucocytes in a unit volume of blood. The normal number is 4,500 to 11,000 WBCs per microliter.\\    
Neutrophil Percentage             & Mature WBC in the granulocytic series. A normal result is between 40\% and 60\%.\\
Neutrophilia                       & Relative neutrophilic leucocytosis, often a result of a bacterial infection.\\
RBC Count                           & Number of erythrocytes in a unit volume of blood. The standard value for children is between 4.0 million to 5.5 million red blood cells per microliter of blood.\\
Hemoglobin                          & Hemoglobin level. Standard value for children is between 10 g/dL and 15 g/dL.\\
RDW                                 & A blood test that measures the differences in the volume and size of the erythrocytes. The normal RDW in children is between 12.3 and 14.1.\\
Thrombocyte Count  & Number of platelets in a unit volume of blood. In healthy pediatric subjects normal count ranges between 250,000 $\mu$L and 450,000 $\mu$L.\\     
CRP                         & Protein produced by the liver. Standard values are no higher than 10 mg/L.\\       
Dysuria                & Pain or other difficulty during urination.\\            
Stool       & Characteristics of bowel movements.\\                  
Peritonitis     & Spasm of abdominal wall muscles detected on palpation.\\                    
Psoas Sign                          & Abdominal pain produced by extension of the hip.\\
Free Fluids                        & Free fluids inside the abdomen.\\
\hline
    \end{tabular}
    \caption{Variables description.}
    \label{descr_var}
\end{table} To construct a binary incidence matrix (and therefore the bipartite network), as described in Section \ref{sec:3}, quantitative and categorical variables are transformed into a suitable number of dummy variables, in a coherent manner with respect to the interpretation provided in Table \ref{descr_var}. In detail, binarization is carried out as follows:
\begin{itemize}
    \item Alvarado Score: value 1 is assigned if the score is greater than 3, which indicates possible/certain appendicitis;
    \item Pediatric Appendicitis Score: value 1 is assigned if the score is greater than or equal to 6, which indicates a possible appendicitis case;
    \item Body Temperature: value 1 is assigned if the body temperature is greater than 37.36;
    \item WBC Count: value 1 is assigned if the count is greater than 11 or lower than 4.5, which corresponds to anomalous values;
    \item Neutrophil Percentage: value 1 is assigned if the percentage is greater than 60 or lower than 40, which corresponds to anomalous values;
    \item RBC Count: value 1 is assigned if the count is greater than 5.5 or lower than 4, which corresponds to anomalous values;
    \item Hemoglobin: value 1 is assigned if the value is greater than 15 or lower than 10, which corresponds to anomalous values;
    \item RDW: value 1 is assigned if the value is greater than 14.1 or lower than 12.3, which corresponds to anomalous values;
    \item Thrombocyte Count: value 1 is assigned if the count is greater than 450 or lower than 250, which corresponds to anomalous values;
    \item CRP: value 1 is assigned if the value is greater than 10, which corresponds to anomalous values;
    \item Stool: value 1 is assigned if the patient has no diarrhea  or constipation.
\end{itemize}
A connection between pair of nodes does exist if the clinical condition is manifested by the patient. The resulting incidence matrix is represented in Figure \ref{mat_data}. 
\begin{figure}[ht!]
    \centering
    \includegraphics[scale=0.5]{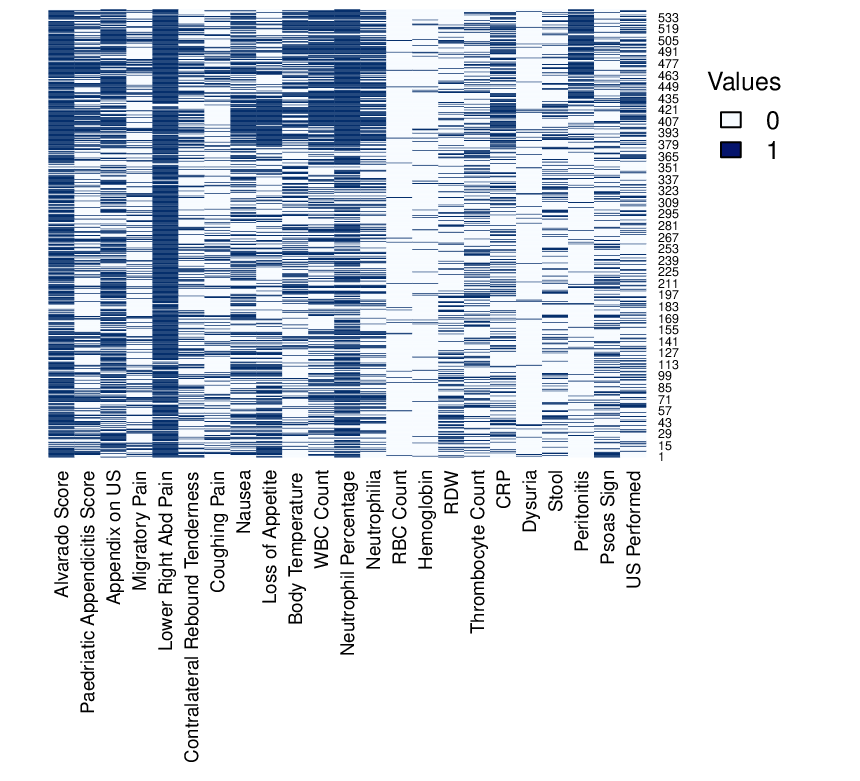}
    \caption{Incidence matrix.}
    \label{mat_data}
\end{figure}\\For a deeper understanding, we report in Figure \ref{var} the distribution of clinical conditions across patients. From this figure it is evident that most patients have detectable vermiform appendix during sonographic examination (appendix on US), and show a high Alvarado Score, lower right abdominal pain, and an anomalous neutrophil percentage. In addition, we can see a rather uniform distribution regarding the Pediatric Appendicitis Score, the loss of appetite, the body temperature, the WBC count, and the neutrophilia. On the contrary, it is evident that more than half of the patients have no migratory pain, contralateral rebound tenderness, coughing pain, abdominal pain produced by extension of the hip (Psoas Sign), dysuria, gastrointestinal problems, peritonitis, and free fluids inside the abdomen. In addition, most patients have normal RBC count, hemoglobin, RDW, Thrombocyte Count, and CRP values.
\begin{figure}[ht!]
    \centering
    \includegraphics[scale=0.4]{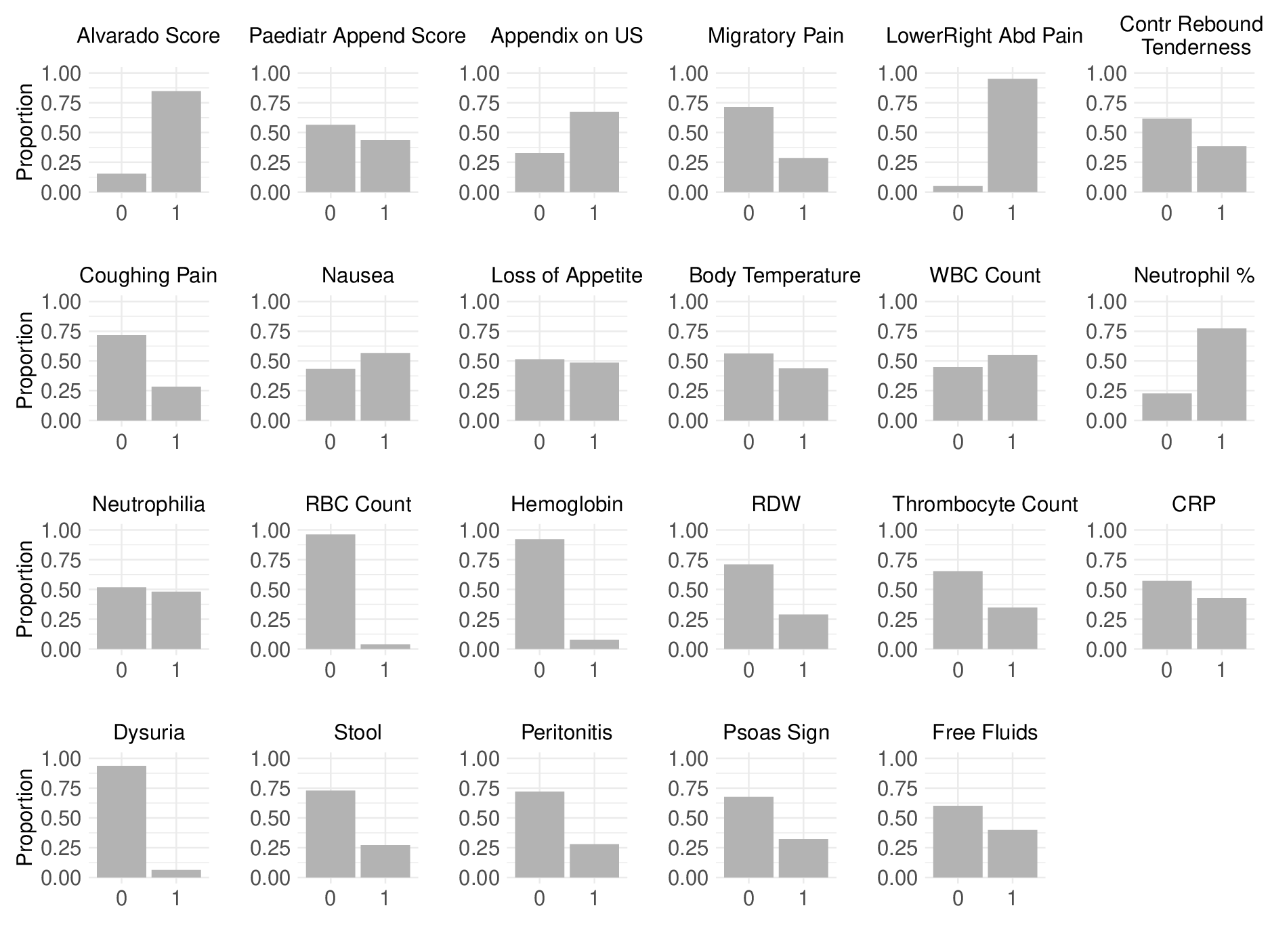}
    \caption{Variables distribution.}
    \label{var}
\end{figure}
As regards patients' characteristics, the following covariates are considered: age, BMI (patient's weight divided by the square of the height), sex, height, weight, length of the stay in the hospital, management (management of the patient assigned by a senior pediatric surgeon - conservative, primary surgical, or secondary surgical), severity (uncomplicated, complicated). A description of such covariates is shown in Figure \ref{cov}. From this Figure, we may see that patients admitted to the hospital are mostly over 10 years old, and with a BMI centered around 15. As regards patient's height and weight, the averages are of about 149 centimeters and 44 kilos, respectively. In addition, the distribution of patients in terms of gender is homogeneous and almost all admissions lasted less than 4 days. Finally, it can be noted that most patients were not advised to have surgery and were diagnosed with low symptom severity.
\begin{figure}
    \centering
    \includegraphics[scale=0.4]{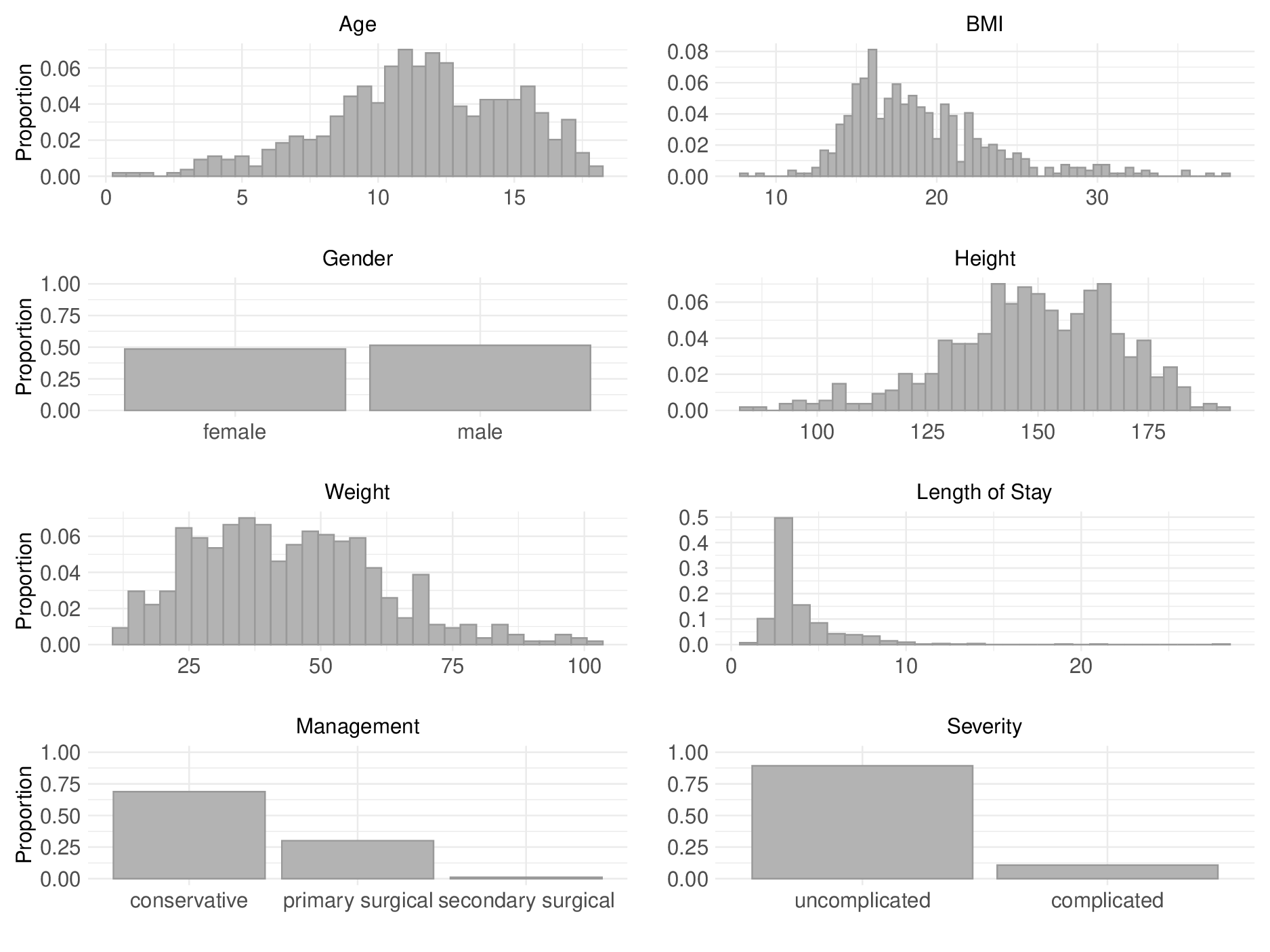}
    \caption{Covariates' distribution.}
    \label{cov}
\end{figure}

\subsection{Results of the proposed biclustering approach}\label{analysis}
The goal is to analyze the bipartite network described so far via the proposed modeling approach to obtain a joint clustering of patients and clinical conditions. More specifically, we aim at identifying groups of patients sharing subsets of clinical conditions, also taking into account how patients' features influence the probability of belonging to a particular component. The proposed model is estimated for a number of components, $G$, and segments, $D$, both ranging from 1 to 4. To prevent the estimation algorithm from remaining trapped in local maxima, a multi-start strategy based on 100 random starts is considered, as detailed in Section \ref{init}. As regards the covariates affecting the latent layer of the model, those described in Section \ref{description} are considered for the specification of $\eta(\boldsymbol{x}_i;\boldsymbol{\beta}_g)$.\\Table \ref{tab:bic} reports the BIC values for all the estimated models; the optimal specification is the one providing the smallest BIC and corresponds to $G = 2$ components and $D=3$ segments.
\begin{table}[ht!]
    \centering
\caption{BIC for varying numbers of components $G$ and segments $D$.}
    \begin{tabular}{l| c c c c}
    \noalign{\smallskip}\hline\noalign{\smallskip}
    \backslashbox{$G$}{$D$} & $1$ & $2$ & $3$ & $4$\\
     \noalign{\smallskip}\hline\noalign{\smallskip}
$1$ & 17096.48 & 15309.17 & 14805.60 & 14895.72  \\
$2$ & 17251.51 & 15053.07 & \textbf{14690.77} & 14810.81   \\ 
$3$ & 17444.27 & 15144.50 & 14954.31 & 15183.33   \\
$4$ & 17658.37 & 15205.96 & 15375.57 & 15704.42 \\
\noalign{\smallskip}\hline\noalign{\smallskip}
    \end{tabular}
    \label{tab:bic}
\end{table}\\Estimates for the vector of parameters $\boldsymbol{b}=[b_1,\dots,b_{G}]$ and $\boldsymbol{\mu}=[\mu_1,\dots,\mu_{D}]$ corresponding to the optimal solution are $\hat{\boldsymbol{b}}$=[-0.59, 0.73] and $\hat{\boldsymbol{\mu}}$=[-2.08, -0.18, 1.69], respectively. These results reveal the existence of two different groups of patients characterized by a more (respectively, less) pronounced symptomatology, together with three groups of clinical conditions, being different in the way they are widespread among patients. In detail, the first, the second, and the third segment group together those conditions that are seldom, mildly, and frequently manifested by the patients admitted to the hospital, respectively. Overall, the model effectively identifies six homogeneous blocks. A clearer picture of these blocks can be obtained by looking at the distribution of the estimated probabilities $\hat{\pi}_{gk}(\hat{\boldsymbol{u}}_i)$ reported in Figure \ref{pred}. These are derived by substituting into Equation (\ref{pi3}) the estimates of $\boldsymbol{b}$, $\boldsymbol{\mu}$, $\boldsymbol{a}_{gk}$, for $g=1, 2$ and $k=1,\dots,23$, and $\boldsymbol{u}_i$. Note that $\hat{\boldsymbol{u}}_i=\text{E}(\boldsymbol{u}_i\mid\boldsymbol{y}_i)$. Looking at these distributions, we may recognize that patients in the first component have a low chance of manifesting clinical conditions classified in segment 1 (mean=0.09) with a reduced variability around this center. On the other side, for these patients, the chance of presenting symptoms belonging to segment 2 and 3 is medium-low (mean=0.35) and medium-high (mean=0.72), respectively. Furthermore, heterogeneity between patients is here more pronounced than in segment 1. When focusing the attention on patients classified in the second component, we may observe that they have a medium-low (mean=0.24), medium-high (mean=0.62), and high (mean=0.89) chance of manifesting clinical conditions belonging to segment 1, 2, and 3, respectively. As far as the heterogeneity captured by the latent trait, we may see from Figure \ref{pred} that this has a low impact on those symptoms belonging to the latter segment.
\begin{figure}[ht!]
    \centering
    \includegraphics[scale=0.5]{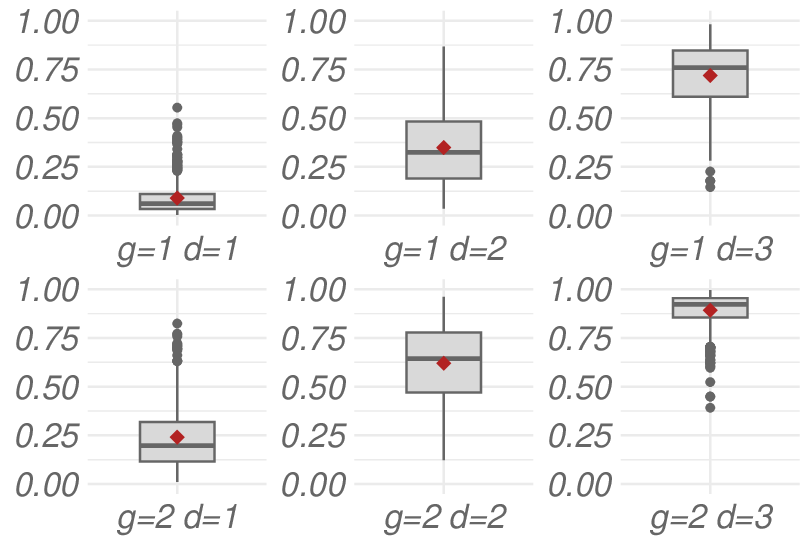}
    \caption{Distribution of the estimated probabilities $\hat{\pi}_{gk}$ for $g=1,\dots,2$ and $d=1,\dots,3$.}
    \label{pred}
\end{figure}\\Table \ref{class} shows the classification of clinical conditions across components. From this table, we may notice that those conditions traditionally attributable to appendicitis (Alvarado Score, Appendix on US, Lower Right Abdominal Pain, and Neutrophil Percentage) are classified in segment 3, regardless the patient component membership. That is, they are generally more likely to be manifested. Other clinical conditions potentially attributable to appendicitis (Pediatric Appendicitis Score, Nausea, Loss of Appetite, Body Temperature, WBC Count, Neutrophilia, CRP, Peritonitis, and Free Fluids) are classified in the second segment, while other more general symptoms (Dysuria, Stool, RBC Count, Hemoglobin, RDW, Thrombocyte Count, Migratory Pain, Contralateral Rebound Tenderness, Coughing Pain, Psoas Sign) are classified in the first segment for patients presenting more severe symptoms (second component). On the other side, for patients belonging to component 1 and manifesting a milder symptomatology, general symptoms such as Body Temperature, Neutrophilia, RBC Count, Hemoglobin, CRP, Dysuria, and Peritonitis, are all classified in the first segment and are therefore rather unlikely to be observed.
\begin{table}[ht!]
    \centering
       \begin{tabular}{l r r}
\hline
  & Component 1 & Component 2\\
\hline
Alvarado Score & 3 & 3\\
\hline
Pediatric Appendicitis Score & 2 & 2\\
\hline
Appendix on US & 3 & 3\\
\hline
Migratory Pain & 2 & 1\\
\hline
Lower Right Abd Pain & 3 & 3\\
\hline
Contralateral Rebound Tenderness & 2 & 1\\
\hline
Coughing Pain & 2 & 1\\
\hline
Nausea & 2 & 2\\
\hline
Loss of Appetite & 2 & 2\\
\hline
Body Temperature & 1 & 2\\
\hline
WBC Count & 2 & 2\\
\hline
Neutrophil Percentage & 3 & 3\\
\hline
Neutrophilia & 1 & 2\\
\hline
RBC Count & 1 & 1\\
\hline
Hemoglobin & 1 & 1\\
\hline
RDW & 2 & 1\\
\hline
Thrombocyte Count & 2 & 1\\
\hline
CRP & 1 & 2\\
\hline
Dysuria & 1 & 1\\
\hline
Stool & 2 & 1\\
\hline
Peritonitis & 1 & 2\\
\hline
Psoas Sign & 2 & 1\\
\hline
Free Fluids & 2 & 2\\
\hline
\end{tabular}
    \caption{Clinical conditions' classification.}
    \label{class}
\end{table}\\Regarding the effect of concomitant variables on patients' clustering, Table \ref{tab:beta2} shows the estimated $\boldsymbol{\beta}_2$ coefficients, together with the corresponding 95\% confidence intervals. Note that, in this application, the first component is the reference, meaning that $\boldsymbol{\beta}_1=\boldsymbol{0}$. Therefore, the estimated coefficients measure the influence of patients' characteristics on the probability of belonging to the second (more affected) component with respect to the first one. From Table \ref{tab:beta2}, it is evident that such a probability is higher for males, and for primary and secondary surgical patients. In contrast, the probability decreases with BMI and it is lower for patients with uncomplicated severity. Age, height, weight, and length of stay do not seem to have a statistically significant effect on the probability of component membership.
\begin{table}[ht!]
    \centering
   \begin{tabular}{l r r}
\hline
  & Estimate & Confidence interval\\
\hline
Intercept & 12.87 & (10.56, 15.18)\\
\hline
Age & -0.05 & (-0.20, 0.10)\\
\hline
BMI & -0.26 & (-0.45, -0.07)\\
\hline
Male & 0.93 & (0.88, 0.99)\\
\hline
Height & -0.04 & (-0.13, 0.05)\\
\hline
Weight & 0.06 & (-0.03, 0.14)\\
\hline
Length of Stay & -0.10 & (-0.44, 0.23)\\
\hline
Management primary surgical & 3.97 & (3.48, 4.46)\\
\hline
Management secondary surgical & 4.04 & (1.61, 6.47)\\
\hline
Severity uncomplicated & -4.04 & (-5.95, -2.12)\\
\hline
\end{tabular}
    \caption{$\boldsymbol{\beta}_2$ estimates and 95\% confidence intervals.}
    \label{tab:beta2}
\end{table}\\To conclude, we report in Figures \ref{mat_ord2} and \ref{mat_ord} the ordered data matrix according to sending and receiving nodes' partitions and the predicted probabilities $\hat{\pi}_{gk}(\boldsymbol{u}_i)$, respectively. These figures clearly highlight the effectiveness of the proposal in properly discovering hidden patterns in the data matrix. 
\begin{figure}
\centering
\begin{minipage}[c]{0.45\textwidth}
\centering
    \includegraphics[scale=0.4]{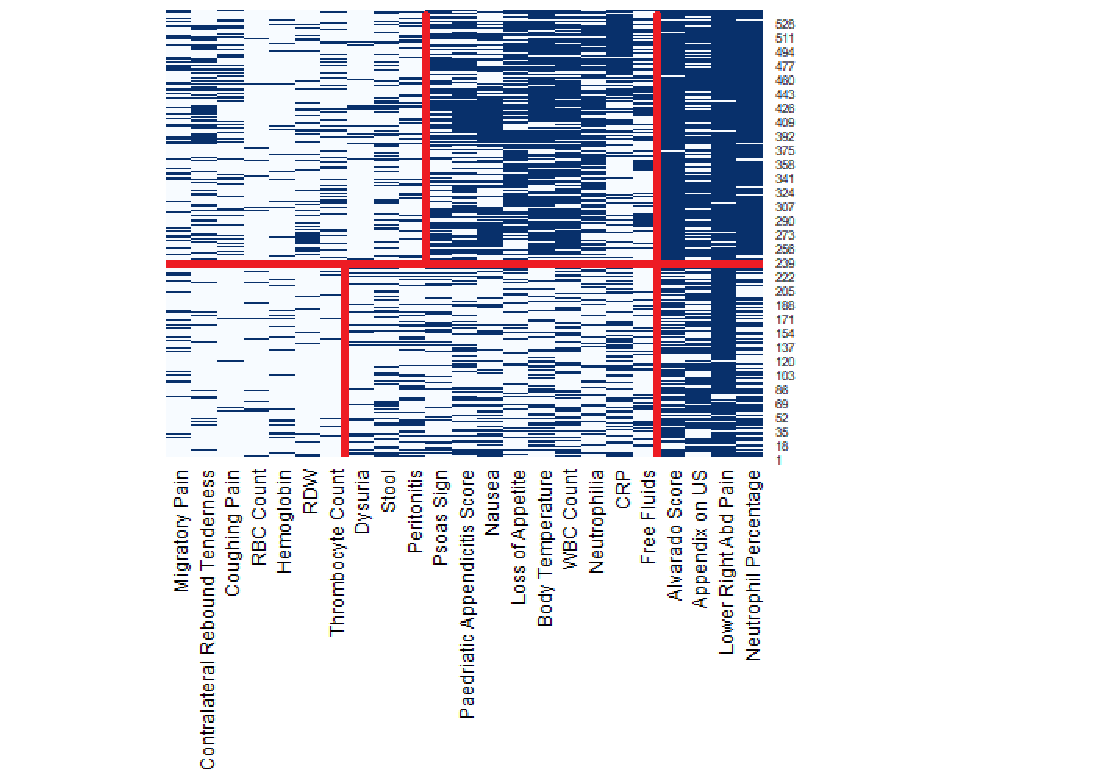}
    \caption{Ordered data matrix.}
    \label{mat_ord2}
\end{minipage}
\begin{minipage}[c]{0.45\textwidth}
\centering
    \includegraphics[scale=0.4]{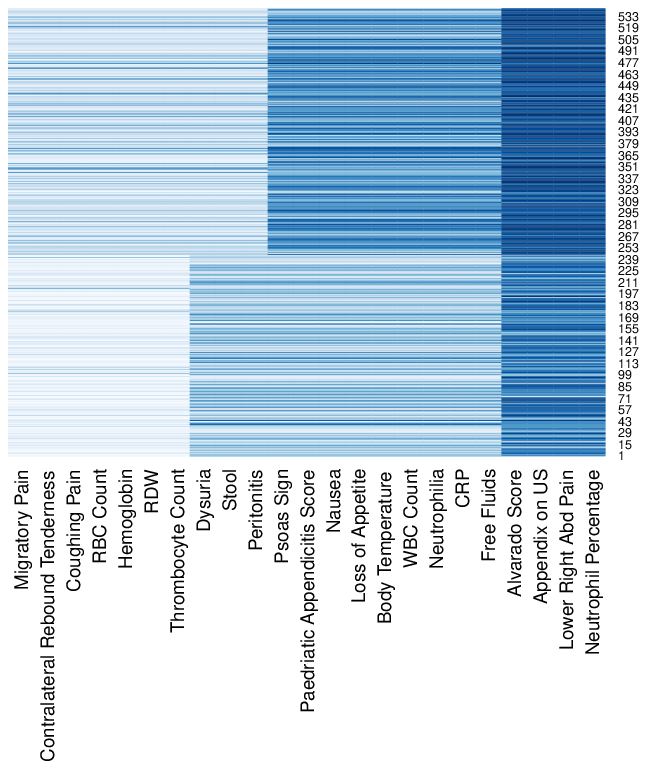}
    \caption{Predicted probabilities $\hat{\pi}_{gk}(\boldsymbol{u}_i)$.}
    \label{mat_ord}
\end{minipage}
\end{figure}

\section{Conclusions}
\label{sec:6}
In this paper we extend the mixture of latent trait analyzers (MLTA) to perform a joint clustering of sending and receiving nodes in a binary bipartite network. In detail, sending nodes are partitioned into clusters called components and, in each of them, receiving nodes are partitioned into clusters called segments. Furthermore, the continuous latent trait considered in the model allows us to capture heterogeneity between sending nodes in the way they connect to receiving nodes. The simulation study we conducted shows that the model can be effectively employed for dimensionality reduction purposes and to identify hidden patterns in a binary data matrix. In detail, when the number of sending and receiving nodes increases, the proposal is able to correctly identify the model parameters and the classification is good. The model is also applied to the bipartite network entailing connections between pediatric patients with suspected appendicitis and their clinical conditions, with the aim of identifying clusters of patients sharing similar subsets of clinical conditions. In detail, the proposed model identifies 2 components of patients according to their symptomatology, and 3 segments of conditions based on their diffusion among patients. Furthermore, the analysis shows that the patients' clustering is influenced by BMI, gender, management, and severity assigned by the doctor. Both the simulation study and the real data analysis have been carried out via the open-source R software \citep{r}. All the codes and the data used in this paper are available upon request.\\
Besides the proposal is detailed for dealing with binary bipartite network, it can be effectively applied also out of the network field, to deal with general binary data matrix. Furthermore, from a methodological point of view, the proposed model can be extended in several directions. A first development may consist in relaxing some restrictions on the latent trait distribution, e.g. by relaxing the unit variance assumption via the inclusion of a parameter that takes variability into account. We may also leave such distribution unspecified and estimate it directly from the data in a non-parametric framework. In addition, the model can also be modified to properly handle response variables with more than two categories. Furthermore, another line of research could involve the analysis of longitudinal bipartite networks, thus allowing for a dynamic joint clustering of sending and receiving nodes, where nodes may move from one partition to another over time.


%


\end{document}